\documentstyle[aps,prl,twocolumn,tighten]{revtex}

\begin{document}

\draft

\preprint{cond-mat/9701141}

\date{Janurary 15, 1997}

\title{Non-perturbative approach to Luttinger's theorem
in one dimension}

\author{Masanori Yamanaka,$^1$ Masaki Oshikawa$^2$ and Ian
Affleck$^{2,3}$}

\address{
Department of Applied Physics,$^1$ University of Tokyo,
Hongo, Bunkyo-ku, Tokyo 113 JAPAN \\
Department of Physics and Astronomy$^2$ and
Canadian Institute for Advanced Research,$^3$
\\ University of British Columbia, Vancouver, BC, V6T 1Z1, CANADA
}

\maketitle

\begin{abstract}
The Lieb-Schultz-Mattis theorem for spin chains is generalized to a wide
range of models of interacting electrons and localized spins in
one-dimensional lattice.
The existence of a low-energy state is generally  proved
except for special commensurate fillings where a gap
may occur. Moreover, the crystal momentum of the constructed low-energy
state is $2k_F$, where $k_F$ is the Fermi momentum of the non-interacting
model, corresponding to Luttinger's theorem. For the Kondo
lattice model, our result implies that $k_F$ must be calculated by
regarding the localized spins as additional electrons.
\end{abstract}

\narrowtext

Strongly correlated electron systems have attracted great interest.
In one spatial dimension (1D), the effect of interactions is often so strong
that the independent electron approximation fails even qualitatively.
The bosonization approach describes a wide range of
one-dimensional interacting electron systems, in which the low-energy
excitations are  better described in terms of bosons rather than
fermions. Such a phase is generally called a Tomonaga-Luttinger
liquid~\cite{TLliquid}.
(For reviews of bosonization, see for example~\cite{boserev}.)
There is an important parameter, the  Fermi momentum $k_F$.
Although  Fermi liquid theory generally breaks down in 1D due to
interactions, the correlation functions still have a singular
wavevector, which is a remnant of the Fermi surface of the free electrons.
Low energy particle-hole like excitations still exist at the
wave-vector $2k_F$. In the free electron model, the Fermi momentum is
determined by the particle density, $\nu$:  $k_F=\pi \nu$ (or $\pi \nu /2$
for spinful electrons in zero magnetic field). Luttinger proved that
interactions do not change the volume inside the Fermi surface, as long as
the system belongs to the Fermi liquid universality
class~\cite{Luttinger:theorem}.
A possible one-dimensional version of
Luttinger's is simply the assertion that gapless neutral excitations exist
at the unrenormalized wave-vector $2k_F = 2 \pi \nu$.  This ought to imply
singularities in Green's functions at the same wave-vector. 
It seems that the absence of a shift in $k_F$ in 1D has been assumed in most
of the literature.

However, since  Fermi liquid theory actually breaks down,
Luttinger's proof does not directly apply to the 1D problems.
A proof for 1D was proposed recently~\cite{Blagoev:Lutt1D},
but only for a very simplified model
with a linearized dispersion relation and without backscattering and
Umklapp term.
An even more difficult example is the Kondo Lattice model, in which
localized spins interact with conduction electrons.
The question arises whether the Fermi momentum is
determined by the density of  conduction electrons only
or by regarding also the localized spins
as electrons.

In this letter, we point out that a simple but powerful theorem
can be applied to a wide range of interacting electron models on 1D
lattice.
A low-energy excited state is explicitly constructed with a definite
crystal momentum, $2 k_F$ where $k_F$ is the free electron Fermi wave-vector
for the specified density, for generic values of the filling factor.
The theorem gives an exact necessary condition for the presence of an
excitation gap above the groundstate.
The theorem itself is independent of
the bosonization approach and can be applied to phases other than the
Tomonaga-Luttinger liquid.
Furthermore, it is applied to the Kondo Lattice,
which is difficult to analyze with standard theoretical techniques.
It shows that the localized spins must be regarded as additional
electrons in computing $k_F$.  (This is sometimes referred to as a
``large Fermi surface'' but we find this notation especially
inappropriate in
one-dimension where the ``Fermi surface'' is just two points.  It is
perhaps better described as Fermi points corresponding to a large filled
Fermi sea.)

Let us start from the simplest case: interacting spinless fermions.
It is known that spinless fermions on a lattice can acquire an excitation
gap due to interactions, even when there is no band gap in
the absence of the interaction.
Such a gapful phase, which is insulating, is called a Mott
insulator~\cite{Mott:insulator}.
(In this paper, we use the phrase in the sense that the insulating
behavior is due to interactions.)
On the other hand, a model of spinless fermions with
short-range interactions can be mapped to an $S=1/2$ chain
with short-range interactions, by the Jordan-Wigner transformation.
The fermion density (number per site) $\nu$ is related to the
magnetization per site $m$ in the spin chain problem as $\nu = m + 1/2$.
Lieb, Schultz and Mattis proved a theorem on the excitation gap
of $S=1/2$ antiferromagnetic chains~\cite{Lieb}.
The Lieb-Schultz-Mattis (LSM) theorem
was later applied to other half-integer spin~\cite{AffleckLieb}.
While these articles focused on the case without a magnetic field, we
recently pointed out that it can be further applied to
spin chains in a magnetic field,
and discussed its relation to possible plateaus in the magnetization
curve~\cite{s32lett}.
Since interacting spinless fermions are mapped to
the $S=1/2$ chain, the LSM theorem should be applicable to the spinless
fermion problem. In the present context, it reads as follows:
%
{\em
Consider a  Hamiltonian with short-range
hopping, and assume it is translationally invariant and also
conserves the total particle number (i.e. has charge $U(1)$ symmetry)
and parity  or time-reversal.  In a chain of length $L$
with periodic boundary conditions, there is at least one low-energy
($O(1/L)$) state above the ground state, if the fermion number per unit cell
$\nu$ is not an integer. The low-energy state has crystal momentum
$2 \pi \nu$ relative to the ground state.
}
The proof is essentially the same as in spin chains.
Here we consider a single band tight-binding model with nearest neighbour
hopping but general interactions written in terms of the electron density,
$n_i$, for simplicity. Generalizations to other models are
straightforward. Define the ``twist'' operator:
\begin{equation}
  U \equiv \exp{\Big[ 2 \pi i \sum_j \frac{j}{L} n_j \Big]} .
\end{equation}
Let $\vert \psi_0 \rangle$ be the ground state of the system. Since
the interaction terms only involve the local density, which is
invariant under transformation by $U$, then:
\begin{eqnarray}
\lefteqn{     \langle \psi_0 | U^{-1} H U - H | \psi_0 \rangle }
\nonumber \\
&& \hspace{10mm}  = -t\left(e^{i2\pi /L}-1 \right)
\sum_i \langle c^\dagger_i c_{i+1}^{\phantom{\dagger}} \rangle
+ \hbox{c.c.},
\label{eq:eofupsi}
\end{eqnarray}
where $t$ is the hopping coefficient and $c_i$ the electron annihilation
operator.
Taylor expanding in powers of
$1/L$, the term of $O(1)$ vanishes assuming that the
Hamiltonian and groundstate are invariant under either parity or
time reversal.
Hence this quantity is $O(1/L)$.
The same result holds for very general types of
Hamiltonians that conserve charge and either parity or time reversal.
While $U | \psi_0 \rangle$ is generally not an eigenstate of the Hamiltonian,
it proves the existence of at least one
low-energy eigenstate,
if $U | \psi_0 \rangle$ is orthogonal to $| \psi_0 \rangle$.
In order to prove the orthogonality, we use  translation, $T$, by
the unit period, which is related to the total crystal momentum
$P$ by $T=e^{i  P}$. (We set the lattice spacing equal to 1.)
$T$ commutes with $H$ and thus $| \psi_0 \rangle$ is an eigenstate of $T$.
(If the groundstate is degenerate it is still possible to construct a
basis of $T$ eigenstates.) We obtain
$  U^{-1} T U = T e^{2 \pi i \nu} $,
where we used $e^{ 2 \pi i n_1} =1$ and $\nu \equiv \sum_j n_j / L$.
This means that the constructed state $U | \psi_0 \rangle$ has
crystal momentum $2 \pi \nu $ relative to the ground state.
Thus they are orthogonal if $\nu$ is not an integer,
concluding the proof.

When $\nu$ is a rational number $p/q$ where $p$ and $q$ are coprimes,
$| \psi_0 \rangle , U | \psi_0 \rangle, \ldots U^{q-1} | \psi_0 \rangle$
are mutually orthogonal and have low energy of $O(1/L)$.
The existence of these low-lying states suggests either a continuum of
gapless excitation or spontaneously broken discrete translation symmetry.
In the latter case, since the $q$ low-lying states differ in the
eigenvalue
of $T$, they are related to spontaneous breaking of  translation symmetry;
the ground state in the thermodynamic limit will  only be invariant under
translation by $q$ unit cells.
Namely, for the filling $\nu = p/q$, the excitation gap can
open only if accompanied by a $q$-fold
spontaneous breaking of the translation symmetry.
An excitation gap without breaking the translation symmetry
is only possible for integer $\nu$.


We note that the low-energy state $U | \psi_0 \rangle$ appearing
in the LSM theorem has the same total particle
number as  the ground state $| \psi_0 \rangle$.
The state $U | \psi_0 \rangle$ at
crystal momentum $ 2 \pi \nu $ implies that the density-density correlation
function has a singularity at this wave vector.
These results are consistent with discussions
in the literature using the bosonization. (For example,
see~\cite{boserev,Giamarchi:Mott1D}.)
In the standard treatment, one only keeps wave-vector components of
the Fermion fields in the vicinity of $\pm k_F$ (left and right movers),
thus obtaining a relativistic interacting field theory which can then be
bosonized. 
Most treatments implicitly assume that $k_F$ is unchanged by
interactions. Our above argument justifies this assumption.

The LSM theorem gives an exact necessary condition for the presence
of an excitation gap.
In terms of bosonization, the gap is generated by a relevant operator.
The low-energy excitation is described by a single bosonic field
$\varphi$ with the Lagrangian ${\cal L} = 1/2 (\partial_{\mu} \varphi)^2$.
The $\varphi$ field has an angular nature 
$\varphi \sim \varphi + 2 \pi R$ where $R$ is the compactification radius.
By symmetries, potentially relevant operators $\cos{n \varphi / R}$'s,
are only allowed at special values
of $k_F$~\cite{Affleck:LesHouches,Giamarchi:Mott1D}.
For $\nu = p/q$ where $p$ and $q$ are coprimes, the leading allowed
operator is $\cos{q \varphi /R}$, which corresponds to a potential
energy with $q$ minima.
If it is a relevant interaction,
the $\varphi$ field in the groundstate will be locked to
one of the $q$ potential minima, corresponding to a $q$-fold
symmetry breaking.  (It is unlikely to be relevant for large $q$ since it
requires a large $R$.)
Thus,  bosonization gives a result consistent completely
with the LSM theorem, if the Fermi momentum $k_F = \nu \pi $
is unchanged by the interaction.
Again this implies the validity of ``Luttinger's theorem'' in 1D.
The above discussions allows
a gapful state at a fractional filling factors, if accompanied
with a breaking of the translation symmetry.
This phenomenon is quite similar to the fractional quantum Hall
effect~\cite{TaoWu:FQHE} where an incompressible fluid is obtained
at a fractional filling of the Landau level, with a breaking of a
(hidden) symmetry.

Now let us turn to interacting spinful electrons.
We assume the model conserves  total electron number
(charge $U(1)$ symmetry), and also the number of up- and down- spin electrons
separately (which corresponds to the conservation of total $S^z$).  We
also assume parity or time reversal invariance,
and electron hopping to be short-ranged.
Since the electron wavefunction has two components, we can construct
two distinct twist operators as
\begin{equation}
  U_{\sigma}  \equiv  
 \exp{\Big[ 2 \pi i \sum_j \frac{j}{L}  n_{\sigma,j} \Big]},
\end{equation}
where $\sigma = \uparrow, \downarrow$ and
$n_{\sigma,j}$ is the fermion
number operator for spin $\sigma$.
Due to the $U(1)$ symmetries and the parity or time reversal symmetry,
transformation by $U_{\sigma}$
raises the energy only by $O(1/L)$, as in~(\ref{eq:eofupsi}).
Since $U_{\sigma}^{-1} T U_{\sigma} = T e^{2 \pi i \nu_{\sigma}}$,
the constructed states $U_{\sigma} | \psi_0 \rangle$
have crystal momentum $2 \pi \nu_{\sigma}$.
This implies the existence of a low-energy state if
either $\nu_{\uparrow}$ or $\nu_{\downarrow}$ is not integer.
As for spinless fermions, for rational $\nu_{\sigma}$'s
the system may acquire a mass gap if accompanied by
a breaking of the translation symmetry.  Again the low energy state has
crystal momentum $2k_{F \sigma}=2\pi \nu_{\sigma}$.
This gives ``Luttinger's theorem'' in 1D for spinful electrons.
In particular, for
the case of zero magnetic field where we expect
$\nu_{\uparrow}=\nu_{\downarrow}=\nu /2$, we get the usual result:
$2k_F=\pi \nu$.

In the bosonization approach to spinful electrons,
a mass gap may be generated by a relevant
interaction, as in the spinless fermion case.
There are separate spin and charge low-energy excitations, corresponding
to the two boson fields $\phi_s$ and $\phi_c$.
Following the discussions in the spinless case,
we find that operators $\cos{n q \phi_c / R}$ are permitted
when $(k_F^{\uparrow} + k_F^{\downarrow}) / \pi = p/q$, where
$p$ and $q$ are coprimes, 
(Actually that with odd $nq$ is further restricted~\cite{Affleck:Banff}.)
When it is relevant, we expect a gap in the charge sector.
Similarly, only when $(k_F^{\uparrow} - k_F^{\downarrow})  / \pi$ takes
a special value, is a spin gap  possible.
To compare the LSM theorem with  bosonization,
it is convenient to define another set of twist
operators by
\begin{equation}
  U_c \equiv U_{\uparrow} U_{\downarrow},  \;\;\;\;
  U_s \equiv U_{\uparrow} U_{\downarrow}^{-1}  .
\label{eq:defucs}
\end{equation}
Let us define also the electron spin operators $s^{\alpha}_j$ by
\begin{equation}
  s_j^{\alpha} \equiv
 \frac{1}{2} \sum_{\mu \nu}
 c^{\dagger}_{\mu,j} \sigma^{\alpha}_{\mu \nu}
 c_{\nu,j}^{\phantom{\dagger}} ,
\label{eq:spinofe}
\end{equation}
where $\sigma^{\alpha}$ is the Pauli matrix.
They are not affected by the transformation by $U_c$ at all:
$U_c^{-1} s_j^{\alpha} U_c = s_j^{\alpha}$.
On the other hand, $U_s$ can be written in terms of the electron
spin operator as
\begin{equation}
  U_s = \exp{ \Big[ 4 \pi i \sum_j \frac{j}{L} s^z_j \Big] }.
\label{eq:spinUs}
\end{equation}
While the LSM-type argument does not tell us whether or not charge-spin
separation occurs,
it is natural to regard $U_c$ ($U_s$) as creating charge
(spin) excitations, when  charge-spin separation does occur.
Again, $U_c | \psi_0 \rangle$ and $U_s | \psi_0 \rangle$ have
energies of $O(1/L)$ due to the symmetries.
Considering the translation $T$,
it can be shown that $U_c | \psi_0 \rangle$ ($U_s | \psi_0 \rangle$)
is orthogonal to $| \psi_0 \rangle$, if total electron number per unit
cell $\nu = \nu_{\uparrow} + \nu_{\downarrow}$
(twice magnetization per site $2m = \nu_{\uparrow} - \nu_{\downarrow}$)
is not an integer.
Thus, without breaking of the translation symmetry,
the charge gap can open only if $\nu$ is an integer, and
the spin gap can open only if $2m$ is an integer.
For rational values of $\nu$ or $2m$, they may open if
accompanied by a breaking of the translation symmetry.

We note that, in~(\ref{eq:spinUs}), $U_s$ has a twist angle
which is twice that used for pure spin chains~\cite{Lieb,AffleckLieb}.
This is necessary to produce a low-energy state in
an electronic model in which $n_j$ can fluctuate.
Thus the spin gap is allowed without breaking the translation symmetry
for integer $2m$.
This gives an interesting difference between the spin degrees of freedom
in the electronic model and the pure spin chain:
while an excitation gap
in the  Heisenberg antiferromagnetic chain must accompany a breaking of
the translation symmetry,
a spin gap without the symmetry breaking
is possible in the Hubbard model at zero magnetization.
Actually, this is what happens in the attractive Hubbard model.
This phenomenon has been also discussed
in terms of field theory~\cite{Affleck:Banff}:
spin degrees of freedom are ``dimerized'' and thus have a gap.
While the dimerization naively means a breaking of translation symmetry,
the spin field is not a local field in terms of the original fermion
operator. Thus the ``dimerization'' of the spin degrees of freedom can
be hidden by the gapless charge degree of freedom.
Intuitively, such a spin-gap phase may be
described by floating singlet pairs.

When the charge gap opens in addition to the spin gap,
the symmetry breaking can  no longer be hidden.
Thus, we may expect that the spin and charge
gap cannot open simultaneously without breaking the translation
symmetry, at half-filling and zero magnetization in a
generalized (single-band) Hubbard model.
Actually, the LSM theorem gives a proof of this statement.
The argument using the original $U_{\sigma}$ gives a stronger result
than that using $U_c$ and $U_s$.
At half-filling and zero magnetization, while the latter proves nothing,
the former proves the existence of a low-lying state,
though the nature of the excitation is unknown.
Thus, assuming that the system has only the charge and spin degrees of
freedom, either charge or spin gap must vanish
or otherwise the translation
symmetry must be broken.

The LSM theorem can be applied to a wide range of models, including
ladders, spin-Peierls system, Kondo lattice, periodic Anderson model, etc.,
with or without orbital degeneracy and/or interactions among conduction
electrons.
As an illustration, we consider the standard Kondo-Heisenberg lattice defined
by the Hamiltonian
\begin{eqnarray}
   H &=& - t \sum_{j, \sigma}
      \left( c^{\dagger}_{\sigma, j+1} c_{\sigma, j}^{\phantom{\dagger}}
      +  \mbox{h.c.}
     \right)
\nonumber \\
  &&  + J_K \sum_j \vec{s}_j \cdot \vec{S}_j
      + J_H \sum_j \vec{S}_j \cdot \vec{S}_{j+1},
\end{eqnarray}
where $\vec{s}_j$ is the electron spin operator defined
in~(\ref{eq:spinofe})
and $\vec{S}_j$ is the localized spin operator of spin $1/2$.
$J_K$ is the Kondo coupling, and $J_H$ is a direct coupling between
the localized spins, which is sometimes introduced in the literature.
The LSM theorem for the Kondo Lattice can be derived using
the twist operators
\begin{equation}
  U_{\pm}  \equiv  \exp{\Big[ 2 \pi i \sum_j
             \frac{j}{L}  ( n_{\pm,j} \pm S^z_j ) \Big]},
\end{equation}
where $\pm$ represents the electron spin.
For these operators,
$  U_{\pm}^{-1} T U_{\pm} = T e^{  \pi i ( \nu \pm 2m + 1)}$
where $\nu$ is the conduction electron
number per site, and $m$ is the magnetization per unit cell
including both the conduction electron and the localized spin.
The identity $e^{2\pi iS^z_1}=-1$ for half-integer localized spins
has been used, as in the LSM theorem for spin-chains.~\cite{Lieb,AffleckLieb}
Charge/spin twist operators can also
be defined as in~(\ref{eq:defucs}) to discuss the conditions for
a charge/spin gap.

The Kondo lattice model exhibits a rich phase
structure~\cite{Tsunetsugu:KLMphase,Zachar,WhiteAffleck,Sikkema:PhD,Shibata:KLFS2}.
In some parameter regions, the low-energy properties of the Kondo lattice
may be described by a Tomonaga-Luttinger liquid or by a
phase with a spin gap but no charge gap.
However, the
Kondo lattice is difficult to treat by bosonization due to the presence
of the localized spins.
While it is possible to introduce the direct exchange
$J_H$ and treat $J_K$
perturbatively~\cite{FujimotoKawakami:KLM,WhiteAffleck,Sikkema:PhD},
it is not clear
whether this approach gives a correct picture for $J_H / J_K \sim 0$.
There have been some
discussions~\cite{Fazekas:KLM,Tsunetsugu:KLMphase,Ueda:LargeFS,FujimotoKawakami:KLM,Moukouri:KLFS,Shibata:KLFS,Shibata:KLFS2}
concerning whether the Fermi wavevector corresponds to a large filled
Fermi sea (including also the localized spins as electrons)
or a small one (including only conduction electrons).
In the large positive $J_K$ limit, all conduction electrons form singlets
with localized spins if $\nu < 1$.
The density of unpaired localized spins, which would behave as
free holes, is $1-\nu$~\cite{Tsunetsugu:KLMphase,Sikkema:PhD},
which leads to the large filled Fermi sea picture.
However, it is questionable whether it applies away from the
large Kondo coupling limit.
On the other hand, we may associate the
Kondo lattice model with an underlying free electron model (the $U\to 0$
limit of the corresponding Anderson model)  with two sites per unit cell
and hence 2 bands.  From a band theory viewpoint, assuming the
bands are nonoverlapping, the Fermi
wavenumber is $\pi (1+\nu )/2$ in the lower band.
If we assume that the wavevector is unaffected by the interactions,
we arrive at the large filled Fermi sea picture.
However, it has been questioned whether the Kondo lattice model
with purely localized spins can be understood correctly in this approach.
While there are several numerical
calculations~\cite{Tsunetsugu:KLMphase,TroyerWurtz:KLM,Ueda:LargeFS,Moukouri:KLFS,Shibata:KLFS,Shibata:KLFS2},
a definite conclusion for whole parameter ranges had not yet been obtained.

In our approach,
the low-energy state appearing in $U_{\sigma} | \psi_0 \rangle$ has
crystal momentum $\pi (\nu \pm 2m +1) $.
In particular, the low-energy state has the momentum
$\pm \pi ( \nu +1)$ (or equivalently $\pm \pi (1-\nu)$) at zero
magnetic field.
This corresponds to a Fermi momentum
 $k_F = \pi ( \nu + 1) / 2=\pi \nu_T/2$, where $\nu_T \equiv \nu +1$, is
the total electron density including the localized spins.
This is consistent with the large filled Fermi sea picture, but
not with the small one.
Thus, our application of the LSM theorem
shows that the large filled Fermi sea picture is
exact for any non-vanishing Kondo coupling $J_K$, with or without
the direct Heisenberg exchange $J_H$.
This result has 
interesting implications for correlation functions in the Kondo Lattice
model.

It is a pleasure to thank Naoto Nagaosa, Subir Sachdev, Arnold E. Sikkema
and Kazuo Ueda for useful discussions.
This work is partially supported by NSERC of Canada.
M.~Y. and M.~O. are supported by JSPS and UBC Killam fellowship,
respectively.

{\it Note Added:}\ 
In the original manuscript, we noted an incompleteness of our proof:
the constructed low-energy state is spread over the entire lattice, and 
it could, in principle,
be merely a pathological excitation which is unphysical in the
thermodynamic limit.
Now Hal Tasaki~\cite{Hal:priv}
has improved our argument so that a low-energy state
is constructed with compact support, thus making the proof complete.
Its details will be reported elsewhere.
We thank Hal Tasaki for the important contribution and for kindly
allowing us to mention his result in the present Letter. 



\end{document}